\documentclass{article}

\usepackage{arxiv}

\usepackage[utf8]{inputenc} 
\usepackage[T1]{fontenc}    
\usepackage{hyperref}       
\usepackage{url}            
\usepackage{booktabs}       
\usepackage{amsfonts}       
\usepackage{nicefrac}       
\usepackage{microtype}      
\usepackage{lipsum}		
\usepackage{graphicx}
\usepackage{natbib}
\usepackage{doi}
\usepackage{amsmath}
\usepackage{algpseudocode}
\usepackage{algorithm}

\title{Compound Auxiliary Metropolis: Incorporating Auxiliary Variables into Multi-Candidate MCMC}

\author{ \href{https://orcid.org/0000-0002-5691-0519}{\includegraphics[scale=0.06]{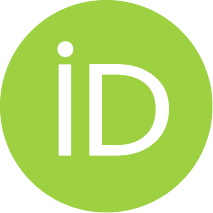}\hspace{1mm}Renny Doig} \\
	Department of Mathematics\\
	Simon Fraser University\\
	Burnaby, BC \\
	\texttt{rennyd@sfu.ca} \\
	\And
	\href{https://orcid.org/0000-0002-8509-7985}{\includegraphics[scale=0.06]{orcid.pdf}\hspace{1mm}Liangliang Wang} \\
	Department of Statistics and Actuarial Science\\
	Simon Fraser University\\
	Burnaby, BC \\
	\texttt{liangliang\_wang@sfu.ca} \\
}

\renewcommand{\shorttitle}{Compound auxiliary Metropolis}

\hypersetup{
pdftitle={Compound auxiliary Metropolis},
pdfsubject={q-bio.NC, q-bio.QM},
pdfauthor={Renny Doig, Liangliang Wang},
pdfkeywords={First keyword, Second keyword, More},
}

\begin{document}

\maketitle

\begin{abstract}
Multiple-try Metropolis (MTM) is a Markov chain Monte Carlo (MCMC) algorithm that improves local transition efficiency by evaluating multiple candidate draws at each iteration. However, for complicated target distributions exhibiting severely non-Gaussian topography or multiple well-separated modes, locally optimal transitions may be insufficient for effective global exploration. In this work, we propose compound auxiliary Metropolis (CAM), a general multi-candidate MCMC method that incorporates both the local state of the chain and auxiliary information into the multi-candidate framework of MTM. Using an auxiliary generating distribution, CAM accommodates a flexible definition of auxiliary information. As examples, we consider three different auxiliary variables: one that promotes state-independent exploration and two that use a reference distribution to improve mixing. These auxiliaries are tested against distributions that present challenging targets for modern MCMC methods. In particular, we focus on the challenges presented by multiple well-separated modes and topography that requires long mixing for local MCMC moves. We find that CAM is able to sample effectively from these distributions, using MTM as a baseline to evaluate the benefit introduced by the auxiliary information. CAM also compares favourably with the No-U-Turn Sampler, showing similar performance for milder test distributions and better performance for the most difficult settings.
\end{abstract}


\section{Introduction}
\label{sec:intro}

Monte Carlo sampling is a fundamental tool of Bayesian inference, particularly when posterior distributions are known only up to an intractable constant of proportionality. To sample from these complex targets, the multiple-try Metropolis (MTM) algorithm \citep{liu2000mtm} extends the standard Metropolis-Hastings (MH) algorithm \citep{metropolis1953,hastings1970} by forming a compound proposal based on multiple candidate draws. Instead of generating and evaluating a single proposal at each iteration, the algorithm draws a pool of candidates, selects one, and evaluates its acceptance using a modified probability that corrects for the multi-candidate selection process. 
By considering multiple candidates at each iteration, MTM increases the probability of making a good transition. This multiple candidate framework improves local exploration and has demonstrated particular success on target distributions exhibiting severely non-Gaussian topography \citep{yang2019componentwise,fontaine2022adaptive,doig2025unified}.

However, optimal local performance does not guarantee effective global exploration. Standard MTM frequently struggles to navigate target distributions with multiple well-separated modes \citep{doig2025unified}. Existing MTM extensions modifying proposal configurations \citep{casarin2013interactingMTM,yang2019componentwise} or candidate weight functions \citep{liu2000mtm,casarin2013interactingMTM,yang2019componentwise,pandolfi2014genMTMpap,gagnon2023localbalancing} primarily refine local mixing rather than resolving issues of global convergence. While ensemble and population-based methods \citep{liang2001real, goodman2010ensemble} address global exploration by simulating multiple interacting chains, they introduce significant computational overhead.

To navigate complex posteriors more efficiently, practitioners often rely on tractable reference distributions or auxiliary variables. 
Reference distributions are chosen for convenience or their similarity to the target distribution. By using a reference to improve initialization or proposals, an algorithm can accelerate burn-in or increase sampling efficiency. Variational methods, for example, yield fast parametric approximations that can serve as reference distributions \citep{gu2015neuralSMC,salimans2019MCMCvariationalgap,surjanovic2022variationalreference}. Similarly, annealing-based methods use a sequence of tempered distributions to smoothly transfer information between the reference distribution and the target. Popular methods that use these tempered paths include annealed importance sampling \citep{neal2001annealedIS}, annealed sequential Monte Carlo (ASMC) \citep{delmoral2006SMCsampler}, and parallel tempering \citep{surjanovic2022variationalreference}. Another commonly used technique in the statistical computation literature is the auxiliary variable. This is some additional variate, {\it i.e.}, not a realization of the target distribution, that is used in the sampling process, often to guide proposals or reduce Monte Carlo variance. An auxiliary variable can take many forms, such as those related to dynamics and adaptation \citep{neal2003slice,hoffman2014nuts,bironlattes2024automala}, indexing multiple distributions \citep{marinari1992simulatedtempering}, data augmentation \citep{tanner1987dataaugmentation}, or those designed for specific inference problems \citep{edwards1988FKSWIsing,green1992SW,higdon1998auxvarbinary}.

In this work, we bridge these methods by proposing Compound Auxiliary Metropolis (CAM), a novel MCMC algorithm that incorporates multiple auxiliary variables directly into the compound proposal step of the MTM framework. By proposing from distributions conditioned on both the current state of the chain and the auxiliary variables, a single CAM iteration can more efficiently explore the state space than the strictly local proposals of MTM. The mechanism by which we incorporate the auxiliary variables is flexible, permitting a wide range of algorithm designs. We provide two examples: an auxiliary variable designed to encourage state-independent global exploration and one that exploits reference distributions (such as those generated via ASMC) to improve mixing. To guarantee detailed balance and establish the theoretical validity of the transition kernel, we frame CAM within the involutive MCMC framework \citep{neklyudov2020involutivemcmc}.

We empirically demonstrate CAM's performance in a simulation study, exploring the effect of different auxiliary variables and other aspects of algorithm design. Two artificial target distributions are used as benchmark distributions, one with multiple well-separated modes and the other with severely non-Gaussian topography. We also evaluate performance on the eight schools problem, a standard target distribution for evaluating MCMC methods. CAM is then compared to standard MH, MTM, and the No-U-Turn Sampler (NUTS) \citep{hoffman2014nuts} as implemented in Stan \citep{stan2024}, a competitive and well-maintained software package used as a performance benchmark.

The remainder of this paper is organized as follows. Section \ref{sec:CAM algorithm P2} formally defines the CAM algorithm, establishes its theoretical relationship to existing MTM variants, and provides practical examples of auxiliary generating distributions. Section \ref{sec:simulation study} details the simulation experiments, comparing different CAM configurations and benchmarking them against established methods. Section \ref{sec:discussion P2} concludes with a discussion of our findings and avenues for future research. 

\section{Compound Auxiliary Metropolis MCMC}
\label{sec:CAM algorithm P2}

While MTM uses a compound proposal to improve local exploration, it can struggle to traverse large distances across the state space, particularly when the target distribution is multimodal. Independent proposals and reference distributions have been successfully used in other MCMC methods to improve sampling performance, such as parallel tempering \citep{surjanovic2022variationalreference} and importance sampling \citep{delmoral2006SMCsampler}. The CAM framework connects these methods by explicitly incorporating auxiliary variables into the compound proposal step, improving both the flexibility and the overall sampling efficiency of the algorithm.

Suppose we wish to sample from a target distribution $\pi$, defined on state space $\mathcal{X}$, whose density or mass function can be evaluated point-wise up to an unknown normalizing constant. What characterizes CAM is the manner in which we introduce auxiliary information into the transition kernel. At each iteration, $I$ auxiliary variables $\mathbf z_1, \ldots, \mathbf z_I$ are drawn from \textit{auxiliary generating distributions} (AGDs), such that $\mathbf z_i \sim f_i$. Because these variables are auxiliary to the target sample, the AGDs are defined on an auxiliary state space $\mathcal{Z}$. This space can be different than $\mathcal{X}$. Examples of how AGDs are specified and how they can improve global sampling efficiency are given in a subsequent section.

For a Markov chain that is currently at state $\mathbf x$, the CAM transition kernel proceeds as follows. First, draw the auxiliary variables $\mathbf z_{1:I}$ from their respective AGDs. Then, generate $M$ proposal candidates from conditional distributions dependent on both the current state and the auxiliary variables: $\mathbf y_m \sim T_m(\cdot|\mathbf x, \mathbf z_{1:I})$. A single proposal, $\mathbf y = \mathbf y_J$, is selected from the candidate pool with some probability. While this selection probability can, in general, take any form \citep{pandolfi2014genMTMpap,doig2025unified}, we adopt the conventional approach of assigning a weight $u_m(\mathbf y_m, \mathbf x, \mathbf z_{1:I})$ to each candidate and selecting candidate index $J$ with probability proportional to its normalized weight:
\begin{equation}\label{eqn:selection prob P2}
P(J|\mathbf y_{1:M}, \mathbf x, \mathbf z_{1:I}) = \frac{u_J(\mathbf y_J, \mathbf x, \mathbf z_{1:I})}{\sum_{m=1}^M u_m(\mathbf y_m, \mathbf x, \mathbf z_{1:I})}.
\end{equation}

To ensure the transition kernel satisfies the detailed balance condition, $M$ reverse samples must be generated following the candidate selection step. The reverse sample at index $J$ is set to the current state, $\mathbf x_J^* = \mathbf x$. The remaining $M-1$ reverse samples are drawn conditional on the selected proposal: $\mathbf x_m^* \sim T_m(\cdot|\mathbf y, \mathbf z_{1:I})$ for $m \ne J$. Note that the values and ordering of the auxiliary variables remain the same between the initial and reverse sampling. The selected proposal $\mathbf y$ is accepted as the next state in the chain with probability:
\begin{equation}\label{eqn:accept prob P2}a(\mathbf x, \mathbf y) = 1 \land \frac{\pi(\mathbf y)T_J(\mathbf x|\mathbf y, \mathbf z_{1:I})P(J|\mathbf x_{1:M}^*, \mathbf y, \mathbf z_{1:I})}{\pi(\mathbf x)T_J(\mathbf y|\mathbf x, \mathbf z_{1:I})P(J|\mathbf y_{1:M}, \mathbf x, \mathbf z_{1:I})}.\end{equation}
If the proposal is rejected, the chain remains at state $\mathbf x$. The CAM kernel presented here produces a valid transition probability for a Markov chain that preserves $\pi$ as its stationary distribution. This procedure is summarized in pseudocode in Algorithm \ref{algo:CAM P2}. 

\begin{algorithm}[h!]
\caption{\bf{Compound auxiliary Metropolis kernel}}
\label{algo:CAM P2}\begin{algorithmic}[1]
\State {\bfseries Input:} (a) Target distribution $\pi$; (b) current state $\mathbf x$.
\State {\bfseries Settings:} (a) Number of candidates $M$; (b) proposal distributions $\{T_m\}$; (c) auxiliary generating distributions $\{f_i\}$.
\State {\bfseries Output:} Next state in the chain.
\State Generate auxiliary variables $\mathbf z_i \sim f_i$ for $i=1,\ldots,I$.\For{$m \in \{1,\ldots,M\}$}
\State Generate candidate $\mathbf y_m \sim T_m(\cdot|\mathbf x, \mathbf z_{1:I})$.
\State Compute forward weights $u_m(\mathbf y_m, \mathbf x, \mathbf z_{1:I})$.
\EndFor
\State Sample $J \in \{1,\ldots,M\}$ with probability given by Equation \ref{eqn:selection prob P2}.
\State Set the selected proposal $\mathbf y = \mathbf y_J$.
\For{$m \in \{1, 2, \dots, M\}$}
\If{$m=J$}
\State Set $\mathbf x_m^* = \mathbf x$.
\Else
\State Generate reverse samples $\mathbf x_m^* \sim T_m(\cdot|\mathbf y, \mathbf z_{1:I})$.
\EndIf
\State Compute reverse weights $u_m(\mathbf x_m^*, \mathbf y, \mathbf z_{1:I})$.
\EndFor
\State Compute the acceptance probability $a(\mathbf x, \mathbf y)$ based on Equation \ref{eqn:accept prob P2}.
\State Generate $U \sim \text{Unif}(0,1)$.
\If{$U < a(\mathbf x, \mathbf y)$}
\State Return $\mathbf y$.
\Else
\State Return $\mathbf x$.
\EndIf
\end{algorithmic}
\end{algorithm}

\subsection{Proof of Detailed Balance via Involutive MCMC}
\label{sec:CAM proof}

To prove that CAM satisfies the detailed balance condition and has $\pi$ as its stationary distribution, we use involutive MCMC \citep{neklyudov2020involutivemcmc}. This framework operates on an extended joint state space through a deterministic involution. As involutive MCMC has been shown to produce valid MCMC algorithms, showing that CAM is an instance of involutive MCMC is sufficient for showing that CAM satisfies detailed balance.

To construct the extended state space for CAM, we augment the current state $\mathbf x$ with all variables generated during the transition. These include the auxiliary variables $\mathbf z_{i} \sim f_i$ for $i=1, \ldots, I$, the candidates $\mathbf y_{m} \sim T_m(\cdot|\mathbf x, \mathbf z_{1:I})$ for $m=1,\ldots, M$, the selected candidate index $J \sim P(J|\mathbf y_{1:M}, \mathbf x, \mathbf z_{1:I})$ where $J\in \{1, \ldots, M\}$, and the $M-1$ reverse samples $\mathbf x_{m}^* \sim T_m(\cdot|\mathbf y_J, \mathbf z_{1:I})$ for $m \neq J$. For notational convenience, we denote the collective vector of these reverse samples as $\mathbf x_{-J}^* = [\mathbf x_{1}^*, \ldots, \mathbf x_{J-1}^*, \mathbf x_{J+1}^*, \ldots, \mathbf x_{M}^*]$. Taking these together define the complete extended state vector:
\begin{equation*}
    \tilde{\mathbf x} = [\mathbf x, \mathbf y_{1:M}, J, \mathbf x_{-J}^*, \mathbf z_{1:I}].
\end{equation*}

Next we specify the distribution over this extended space. Assuming the chain is in stationarity, the marginal distribution of the current state is $\pi(\mathbf x)$. The joint distribution over the extended space, $\tilde\pi(\tilde{\mathbf x})$, is therefore the product of $\pi(\mathbf x)$ and the conditional generative distributions of the remaining variables:
\begin{equation}\label{eqn:extended dist}
    \tilde\pi(\tilde{\mathbf x}) = \pi(\mathbf x) \left[ \prod_{i=1}^I f_i(\mathbf z_i) \right] \left[ \prod_{m=1}^M T_m(\mathbf y_m|\mathbf x, \mathbf z_{1:I}) \right] P(J|\mathbf y_{1:M}, \mathbf x, \mathbf z_{1:I}) \prod_{m \ne J} T_m(\mathbf x_m^*|\mathbf y_J, \mathbf z_{1:I}).
\end{equation}

To evaluate the acceptance of the selected proposal $\mathbf y_J$, we define a deterministic involution $g(\tilde{\mathbf x})$ that defines how the state changes in the event that the forward transition results in the acceptance of the proposed state. In the case of CAM, the involution swaps the current state $\mathbf x$ with the selected candidate $\mathbf y_J$, and the unused candidates $\mathbf y_{-J}$ with the corresponding reverse samples $\mathbf x_{-J}^*$. The auxiliary variables $\mathbf z_{1:I}$ and the selection index $J$ remain unchanged. The transformation is given by:
\begin{equation*}
    g(\tilde{\mathbf x}) = [\mathbf y_J, \mathbf x_{1:J-1}^*, \mathbf x, \mathbf x_{J+1:M}^*, J, \mathbf y_{-J}, \mathbf z_{1:I}].
\end{equation*}
Because $g$ is a permutation of the components of $\tilde{\mathbf x}$, the absolute value of the Jacobian determinant for its continuous components is exactly 1.

In the involutive MCMC framework, the acceptance probability that satisfies detailed balance reduces to the density ratio between the transformed and original extended states:
\begin{equation*}
    a(\mathbf x, \mathbf y_J) = \min\left\{1, \frac{\tilde\pi(g(\tilde{\mathbf x}))}{\tilde\pi(\tilde{\mathbf x})}\right\}.
\end{equation*}
Substituting the joint distribution from Equation \ref{eqn:extended dist} into this ratio, the terms corresponding to the AGDs $f_i(\mathbf z_i)$ and the unselected candidate and reverse proposals cancel:
\begin{align*}
    & \frac{\tilde\pi(g(\tilde{\mathbf x}))}{\tilde\pi(\tilde{\mathbf x})} \\=& \frac{\pi(\mathbf y_J) \left[ \prod_{i=1}^I f_i(\mathbf z_i) \right] \left[ \prod_{m=1}^M T_m(\mathbf x_m^*|\mathbf y_J, \mathbf z_{1:I}) \right] P(J|\mathbf x_{1:M}^*, \mathbf y_J, \mathbf z_{1:I}) \prod_{m \ne J} T_m(\mathbf y_m|\mathbf x, \mathbf z_{1:I})}{\pi(\mathbf x) \left[ \prod_{i=1}^I f_i(\mathbf z_i) \right] \left[ \prod_{m=1}^M T_m(\mathbf y_m|\mathbf x, \mathbf z_{1:I}) \right] P(J|\mathbf y_{1:M}, \mathbf x, \mathbf z_{1:I}) \prod_{m \ne J} T_m(\mathbf x_m^*|\mathbf y_J, \mathbf z_{1:I})} \\
    = & \frac{\pi(\mathbf y_J)T_J(\mathbf x|\mathbf y_J, \mathbf z_{1:I}) P(J|\mathbf x_{1:M}^*, \mathbf y_J, \mathbf z_{1:I})}{\pi(\mathbf x)T_J(\mathbf y_J|\mathbf x, \mathbf z_{1:I}) P(J|\mathbf y_{1:M}, \mathbf x, \mathbf z_{1:I})}.
\end{align*}
Thus, the involutive MCMC acceptance probability simplifies to:
\begin{equation*}
    a(\mathbf x, \mathbf y_J) = \min\left\{1, \frac{\pi(\mathbf y_J)T_J(\mathbf x|\mathbf y_J, \mathbf z_{1:I}) P(J|\mathbf x_{1:M}^*, \mathbf y_J, \mathbf z_{1:I})}{\pi(\mathbf x)T_J(\mathbf y_J|\mathbf x, \mathbf z_{1:I}) P(J|\mathbf y_{1:M}, \mathbf x, \mathbf z_{1:I})}\right\}.
\end{equation*}
This matches the CAM acceptance probability that was presented in Equation \ref{eqn:accept prob P2}, confirming that the CAM algorithm is a valid, reversible MCMC method that targets the stationary distribution $\pi$.

\subsection{Auxiliary Generating Distributions}
\label{sec:AGDs P2}

The form of the AGD is not restrictive, allowing the user to choose $\mathcal Z$ and $f$ to suit the needs of their particular problem. In this section, we illustrate how these AGDs can be constructed and how the auxiliary variables are incorporated into the compound proposal mechanism. Generally, an AGD can be any valid probability distribution defined on an auxiliary state space $\mathcal Z$. This space need not be identical to the target state space $\mathcal X$, provided the conditional proposal distributions still map back to the target space: $T(\cdot|\mathbf x,\mathbf z_{1:I}) : (\mathcal X \times \mathcal Z^I) \to \mathcal X$. We classify auxiliary variables into two categories: \textit{uninformed} and \textit{informed}. Uninformed auxiliary variables contain no prior information about the structure of the target distribution. Their purpose is to help the chain jump between well-separated regions of high probability in a single transition by making proposals that are independent of the local topology. Conversely, informed auxiliary variables are constructed such that they approximate features of the target distribution, thereby accelerating burn-in and improving stationary sampling efficiency. For simplicity, the remainder of this paper focuses on the case where a single auxiliary variable is used at each iteration ($I=1$).

As an example of an uninformed AGD, we consider a discrete, uniform grid spanning a bounded region of the state space. Let $\mathcal{S} = \{L, L+\Delta, L+2\Delta, \dots, U\}$ define a sequence of evenly spaced points in one dimension between $L$ and $U$. For a $D$-dimensional target, the auxiliary space is the Cartesian product $\mathcal Z = \mathcal{S}^D$. The AGD is then simply a uniform distribution over this grid:
\begin{equation}\label{eqn:discrete AGD P2}
    \mathbf z \sim \text{Uniform}(\mathcal{S}^D).
\end{equation}
While the choice of $L$ and $U$ would contain some information about $\pi$ if $\mathcal X$ does not naturally suggest them, there is no information contained about the topography of $\pi$ within that region. 

To illustrate the construction of an informed AGD, we use annealed sequential Monte Carlo to generate a coarse initial approximation of the target distribution. While a detailed description and pseudocode for ASMC are provided in Section A of the Supplementary Materials, we summarize the algorithm here. ASMC is an importance sampling method that generates a weighted sample from the target by propagating particles through a sequence of artificial distributions. These distributions are defined as $\pi_r \propto \pi^{\alpha_r}\rho^{(1-\alpha_r)}$, where $\rho$ is a tractable reference distribution and $0=\alpha_0 < \alpha_1 < \dots < \alpha_R=1$ is a sequence of annealing parameters \citep{delmoral2006SMCsampler,wang2020ASMCphylo}. This yields a tempered path linking the tractable reference distribution to the intractable target. The reference distribution must have the same state space as the target distribution and must be one for which independent samples can be directly drawn. To avoid the difficult task of manually tuning the entire annealing sequence, the parameter $\alpha_r$ can be determined adaptively at each step \citep{zhou2016toward,wang2020ASMCphylo}. The output of ASMC is a collection of $K$ weighted particles for each annealed distribution in the sequence:
\begin{align*}
    &\left\{W_r^{(k)}, \mathbf x_r^{(k)}\right\}_{k=1}^K, \quad \text{yielding the empirical measure} \quad \hat\pi_r(\mathbf x) = \sum_{k=1}^K W_r^{(k)}\delta_{\mathbf x_r^{(k)}}(\mathbf x),
\end{align*}
where $W_r^{(k)}$ is the normalized weight of the $k$-th particle from the $r$-th distribution, and $\delta$ denotes the Dirac delta mass. From this ASMC output, we define two informed AGDs. The first uses only the final weighted sample from the target distribution:
\begin{equation}\label{eqn:asmc-post AGD}
    \mathbf z \sim \hat\pi_R.
\end{equation}
The second incorporates information from the entire annealing path by treating the AGD as a uniform mixture of all $R+1$ intermediate empirical distributions:
\begin{equation}\label{eqn:asmc-all AGD}
    \mathbf z \sim \frac{1}{R+1}\sum_{r=0}^R \hat\pi_r.
\end{equation}
Each choice has potential merits and demerits. If $\hat\pi_R$ is used as the AGD, then it will be closer to the target than that given in Equation \ref{eqn:asmc-all AGD}; this may produce a more efficient sampler, but only if $\hat\pi_R$ is reasonably close to $\pi$. On the other hand, the AGD using all $R+1$ distributions may be a worse approximation of $\pi$ than $\hat\pi_R$, however by including samples from the full annealing path it may promote better exploration of the state space.

In the preceding discussion of informed AGDs based on ASMC, the ASMC algorithm is executed only once as a pre-processing step prior to running the CAM algorithm. During the evaluation of a single CAM transition, generating an auxiliary variable $\mathbf z$ from $\hat\pi_R$ simply requires drawing a particle $\mathbf x_R^{(k)}$ from a categorical distribution with probabilities proportional to the normalized weights $W_R^{(k)}$. This is a computationally trivial operation that adds little overhead to the MCMC transition. We select ASMC as the foundation for our informed AGDs because its weight updates are highly parallelizable \citep{wang2020ASMCphylo} and it can effectively capture multiple modes simultaneously \citep{wang2021semiparametricSMC}. Moreover, the coarseness of the annealing scheme can be controlled through a tuning parameter, allowing us to quickly compute a sample that serves as a suitable global guide for the auxiliary proposals.

The CAM framework can easily accommodate other computationally inexpensive methods for constructing informed AGDs. For instance, variational inference methods can quickly fit a tractable, parametric approximation to the target which could serve as the AGD. Alternatively, one can compute the maximum likelihood estimate $\hat{\pmb\theta}$ of the parameter(s) of interest and take its estimated 
asymptotic distribution $\mathbf z \sim \mathcal N(\hat{\pmb\theta}, V(\hat{\pmb\theta}))$ as the AGD. Finally, informed auxiliaries offer a novel mechanism for integrating meta-analyses into standard Bayesian workflows; if $n$ independent estimates $\hat{\pmb\theta}_{1:n}$ are collated from existing literature, the AGD can be defined to sample directly from this empirical prior pool.

\subsection{Configuring the Proposal and Selection Mechanisms}
\label{sec:weight functions P2}

As a multi-candidate MCMC algorithm, CAM has two other components, in addition to the number of candidates, that govern its performance: the weight function and the configuration of the proposal distributions. Both of these should be determined by the user based on their particular analysis. As the multi-candidate proposal mechanism in CAM is similar to that of MTM, we borrow from the MTM literature in our discussion of how these components can be specified.

The weight function $u_m(\mathbf y_m, \mathbf x, \mathbf z_{1:I})$ controls the probability of selecting candidate $\mathbf y_m$. For notational convenience and to align with existing MTM literature, we momentarily drop the explicit dependence on the auxiliary variables, writing the weight simply as $u_m(\mathbf y_m, \mathbf x)$; however, in practice, the weight function can depend on $\mathbf z_{1:I}$. While the selection probability can theoretically take any form, normalized weight functions are the standard approach \citep{doig2025unified}. The only requirement placed on the weight function is that $u_m(\mathbf y_m, \mathbf x)$ must be strictly positive on the support of the target distribution. 
Several forms of weight functions can be found in the MTM literature. The original formulation \citep{liu2000mtm} restricts the weight to be of the form $$u_m(\mathbf y_m, \mathbf x) = \pi(\mathbf y_m)T_m(\mathbf x|\mathbf y_m)\lambda(\mathbf y_m, \mathbf x),$$ where the scaling function $\lambda(\mathbf y_m, \mathbf x)$ is positive on the support of $\pi$ and symmetric with respect to $\mathbf y_m$ and $\mathbf x$. One example of this is the ``proportional'' weight function, $u_m(\mathbf y_m, \mathbf x) = \pi(\mathbf y_m)$, which selects the $m$-th candidate with probability proportional to its target density. Another variation is the ``jump distance'' weight function \citep{yang2019componentwise}:
\begin{equation*}
    u_m(\mathbf y_m, \mathbf x) = \pi(\mathbf y_m)\|\mathbf y_m - \mathbf x\|^\alpha,
\end{equation*}
which places greater weight on candidates further from the current state. The tuning parameter $\alpha$ controls this distance penalty, with empirical evidence suggesting an optimal range of $\alpha \in (2,4)$. More recently, the symmetry restriction was relaxed to allow more generally defined weight functions \citep{pandolfi2014genMTMpap}. A notable example is the ``locally balanced'' weight function, $u_m(\mathbf y_m, \mathbf x) = \sqrt{\pi(\mathbf y_m)}$, which has theoretical properties regarding convergence rates and pre-stationary performance \citep{gagnon2023localbalancing}. However, an independent empirical benchmark of MTM algorithms found negligible evidence that the choice of weight function significantly impacts stationary sampling efficiency \citep{doig2025unified}. Consequently, in this work, we implement the locally balanced weight function as it is the only weight function with known properties beyond preserving detailed balance.

The second component is the proposal distribution configuration. Practitioners must determine whether candidates are proposed jointly over the full state vector or updated sequentially in a component-wise fashion. Recent benchmarking indicates that this choice significantly impacts sampling efficiency, with the optimal architecture depending heavily on the target distribution's topography \citep{doig2025unified}. 
Additionally, the incorporation of auxiliary variables introduces new structural design choices. In this work, we partition the $M$ candidates into $M_x$ \textit{local proposals} and $M_z$ \textit{auxiliary proposals} such that $M = M_x + M_z$. The local proposals depend on the joint state strictly through the current position of the chain (i.e., $T_m(\mathbf y_m|\mathbf x, \mathbf z_{1:I}) = T_m(\mathbf y_m|\mathbf x)$). Conversely, the auxiliary proposals condition only on the auxiliary variables (i.e., $T_m(\mathbf y_m|\mathbf x, \mathbf z_{1:I}) = T_m(\mathbf y_m|\mathbf z_{1:I})$), rendering them independent of the chain's current local position. While more complex dependencies can be specified in a manner similar to interacting-chain MTM algorithms \citep{casarin2013interactingMTM}, this partitioned approach simplifies both implementation and analysis.

\subsection{Relationship with Multiple-Try Metropolis}
\label{sec:MTM P2}

CAM is closely related to MTM, the original multi-candidate MCMC algorithm. In this section we show how MTM can be recovered from CAM. We begin with Metropolis-Hastings, as MTM is itself an extension of MH. Setting $M_x = 1$ and $M_z = 0$ recovers the standard random-walk MH algorithm \citep{metropolis1953}; with $M_z = 1$ and $M_x = 0$ we recover the independence-chain MH algorithm \citep{hastings1970}. To recover MTM, that is multi-candidacy without the auxiliary variables, set $M_x> 1$ and $M_z=0$. Alternatively, if one sets $M_x=0$ and $M_z\ge1$, one recovers the independent multiple-try Metropolis algorithm \citep{martino2018review}. A full description of MTM, including pseudocode, is in the Supplementary Materials.

There are also two multi-chain variations of MTM, wherein parallel chains interact during the proposal stage. The first is the interacting-chain MTM (IC-MTM) algorithm \citep{casarin2013interactingMTM}, which runs several MTM chains simultaneously. At each iteration, the $M$ candidates for a given chain are drawn conditional on the current states of the other parallel chains, facilitating information sharing between chains. Because these parallel chains are Markovian, the IC-MTM transition for any of its constituent chains is equivalent to a CAM transition where the current states of the other chains act as the auxiliary variables. Similarly, the annealed IC-MTM algorithm \citep{casarin2013interactingMTM} runs $R+1$ simultaneous Markov chains. Unlike standard IC-MTM, the $r$-th chain targets a tempered distribution $\pi_r \propto \pi^{\alpha_r}\rho^{1-\alpha_r}$, where $\rho$ is a reference distribution and $0=\alpha_0 < \alpha_1 < \cdots < \alpha_R = 1$. The first $R$ chains are updated via standard MH steps, while the $R$-th chain is updated using an MTM step conditioned on the states of the annealed chains. Under the CAM framework, the current states of the chains targeting these $R$ intermediate distributions serve as the auxiliary variables $\mathbf z_{1:I}$.

\section{Simulation Experiments}
\label{sec:simulation study}

To systematically evaluate the performance of CAM, we conduct a series of simulation experiments targeting distributions that are characterized by severe non-Gaussian topography or multimodality. We first compare different CAM configurations against one another to isolate the effects of the AGD and the proposal configuration. Subsequently, the best CAM configuration is evaluated against other MCMC algorithms to provide a reasonable assessment of the proposed algorithm's performance. 

\paragraph*{Novel CAM Algorithms} 
We evaluate six distinct CAM configurations by crossing three AGDs with two proposal configurations. The three AGDs are: (1) the uninformed discrete uniform grid (Equation \ref{eqn:discrete AGD P2}), (2) the target-only ASMC empirical distribution (ASMC-R; Equation \ref{eqn:asmc-post AGD}), and (3) the pooled intermediate ASMC empirical distribution (ASMC-0:R; Equation \ref{eqn:asmc-all AGD}). The two proposal configurations evaluated are full-block (joint) updates and component-wise updates. 
Local random-walk proposals use an adaptive (co)variance scheme: full-block proposals use empirical covariance adaptation \citep{fontaine2022adaptive}, while component-wise proposals use the balanced selection rate scheme \citep{yang2019componentwise} (detailed in the Supplementary Materials). Auxiliary proposals use standard Gaussian distributions with unit (co)variance. Across all settings, an equal number of local and auxiliary candidates are drawn at each iteration ($M_x = M_z = M/2=M'$).

\paragraph*{Other Algorithms} 
The CAM framework is benchmarked against three algorithms: MH, MTM, and the Hamiltonian Monte Carlo (HMC). MTM uses heterogeneous component-wise proposals, balanced selection rate adaptation, and the locally balanced weight function. Similarly, MH uses component-wise proposals with adaptive variances \citep{andrieu2008adaptiveMCMC}. These are included in the study due to their relationship with CAM. HMC is included as a competitive benchmark for modern MCMC performance. We use NUTS, implemented in the probabilistic programming language Stan \citep{stan2024}.

\paragraph*{Execution and Computing Resources} 
The CAM, MTM, and MH algorithms were implemented in Julia \citep{julia}, while Stan was accessed via the Stan.jl library \citep{goedman2012stanjulia}. To ensure a fair comparison of computational efficiency, 50 independent chains were generated for each configuration under a strictly fixed wall-clock time budget. Note that for the CAM configurations using ASMC-based AGDs, the time required to execute the initial ASMC pre-processing step is included within this fixed time budget.

\paragraph*{Post-Processing} 
To automatically determine the burn-in period for each generated sample, we used the average $\hat R$ statistic \citep{vehtari2021ranknormalization} across all coordinates, discarding initial samples until convergence was indicated. Because $\hat R$ is an unreliable diagnostic for multimodal targets \citep{vehtari2021ranknormalization}, burn-in for the Gaussian mixture experiment was handled using a separate, distance-based truncation method described in Section 3.1. Effective sample sizes (ESS) and $\hat R$ diagnostics were computed in Julia using the MCMCChains.jl library \citep{smithmcmcchains}. All visualizations were rendered in R \citep{r} using the ggplot2 package \citep{ggplot2wickham2016}. These experiments can be replicated at the following git repository: \url{https://github.com/SFU-Stat-ML/cam-experiments}.

\subsection{Multivariate Gaussian Mixture}
\label{sec:gaussian_mixture}

To assess how well a single Markov chain of each algorithm can navigate a multimodal target distribution, we use a two-dimensional, five-component Gaussian mixture model. The distance between component means is used to scale sampling difficulty. These component means are located at $[0,0]^T$, $[\delta,\delta]^T$, $[-\delta,-\delta]^T$, $[\delta,-\delta]^T$, and $[-\delta,\delta]^T$, where $\delta$ is used to control the intermodal distance. Each component features a unit covariance matrix, and the mixture weights are strictly unbalanced at $[0.1, 0.2, 0.4, 0.2, 0.1]$. We evaluate performance across increasing intermodal distances, setting $\delta \in \{5, 10, 15, \ldots, 30\}$, and test each CAM configuration across candidate pool sizes $M' \in \{2, 10, 100, 1000, 10000\}$. 

Because standard convergence diagnostics often fail to detect mode-dropping in multimodal landscapes, we evaluate global exploration and proper mixture-weight recovery by computing the two-sample Kolmogorov-Smirnov distance (KSD). The KSD is calculated between the algorithm-generated samples and a baseline sample drawn exactly from the target distribution. For subsequent comparisons against external methods, the optimal combination of $M$ and AGD is selected for both the full-block and component-wise CAM configurations.

\begin{figure}[h!]
\centering
\includegraphics[width=\textwidth]{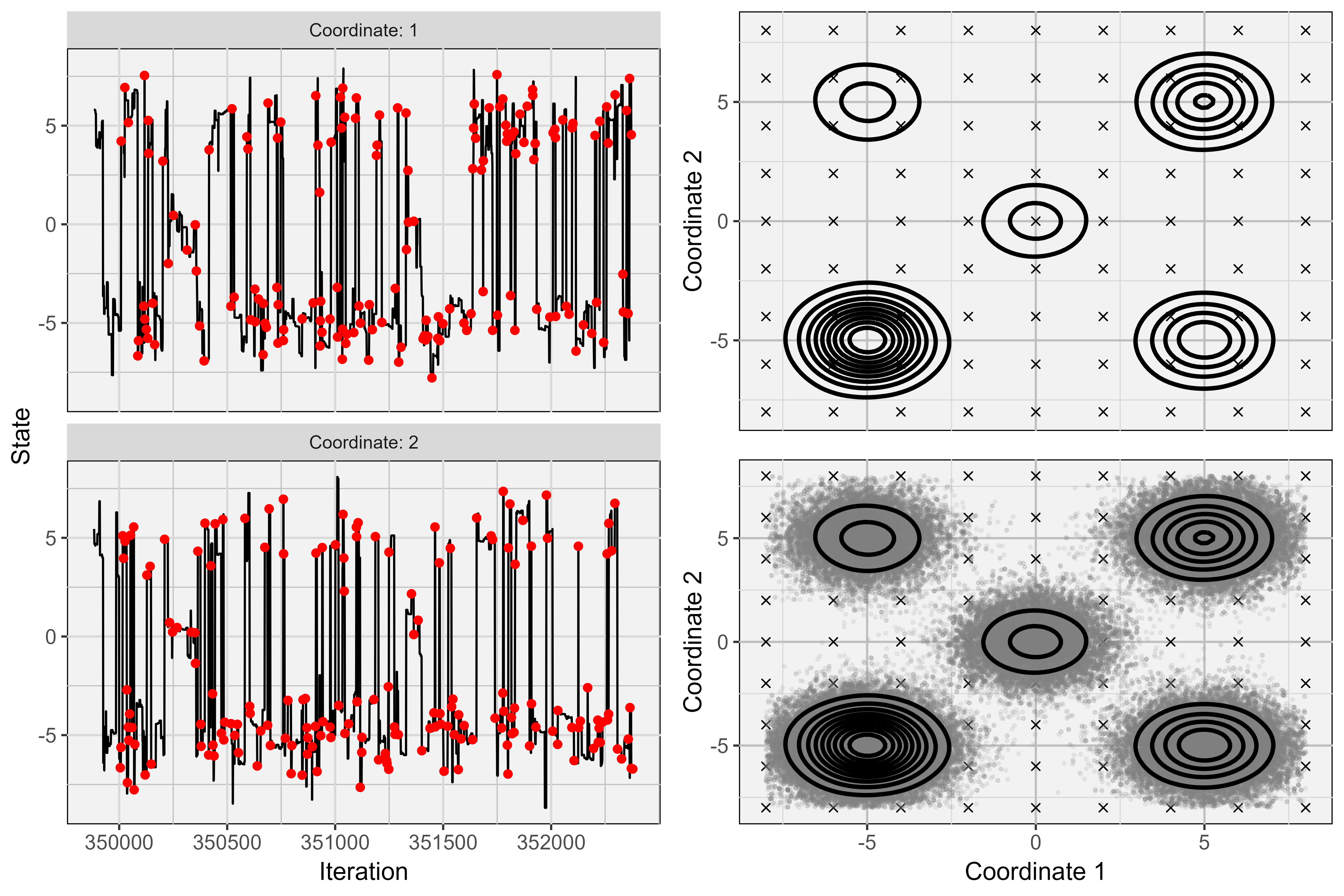}
\caption{CAM with an uninformed discrete auxiliary. Left: Traceplots of the final 2,500 samples for both coordinates with red points indicating transitions sampled from the auxiliary variable. Top-right: Target distribution contours (black) with the discrete auxiliary grid (crosses). Bottom-right: The resulting CAM samples overlaid with grey points.}
\label{fig:discrete traceplot}
\end{figure}

Before analyzing the experiment results, we present a proof-of-concept demonstrating how incorporating auxiliary variables facilitates multimodal exploration. Figure \ref{fig:discrete traceplot} visualizes a CAM run with a discrete uninformed AGD on a mixture with $\delta=5$. The top-right panel overlays the grid of the auxiliary state space onto the contours of the target distribution; the bottom-right panel confirms that the generated chain successfully sampled from all five modes. The traceplots (left panels) highlight transitions where the auxiliary candidate was selected (red points). Importantly, the drastically different geometry of the auxiliary proposal does not adversely affect the resulting sample. Instead, it actively drives both intra-mode mixing and large inter-mode jumps.

\begin{figure}[h!]
\centering
\includegraphics[width=\textwidth]{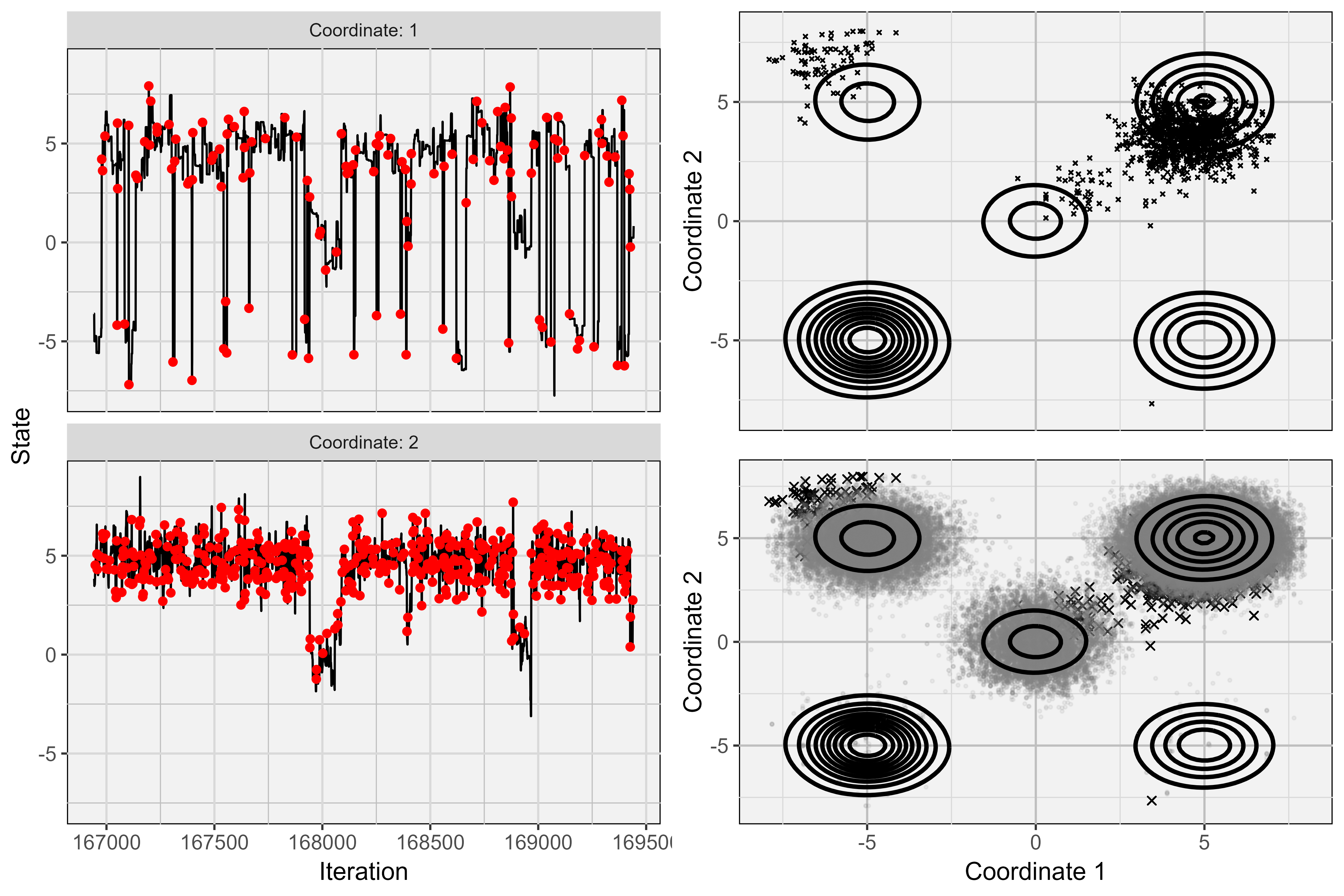}
\caption{CAM with an informed ASMC auxiliary. Left: Traceplots of the final 2,500 samples for both coordinates, with red points indicating auxiliary transitions. Top-right: Target contours (black) with ASMC auxiliary samples (crosses). Bottom-right: The resulting CAM samples overlaid in grey points.}
\label{fig:asmc traceplot}
\end{figure}

The results in Figure \ref{fig:asmc traceplot} are similar to those above, but using the informed ASMC-R AGD. Unlike the discrete grid, the ASMC auxiliary concentrates its mass heavily near the target modes. However, in this specific pre-computation, the ASMC algorithm failed to capture two of the five modes. Consequently, while the CAM chain efficiently explores the three discovered modes, it entirely neglects the remaining two. As with the discrete auxiliary, the traceplots confirm that the informed auxiliary candidates successfully facilitate jumps between the identified modes. 

This proof-of-concept highlights the core utility of the CAM framework: auxiliary proposals enable traversing low-probability regions that would otherwise trap local random-walk kernels. Furthermore, it illustrates a fundamental trade-off in AGD design. While an informed auxiliary (like ASMC) offers highly efficient local targeting, it risks trapping the chain if the initial approximation drops modes. Conversely, an uninformed grid improves global coverage but may yield lower localized acceptance rates.

\begin{figure}[h!]
\centering
\includegraphics[width=\textwidth]{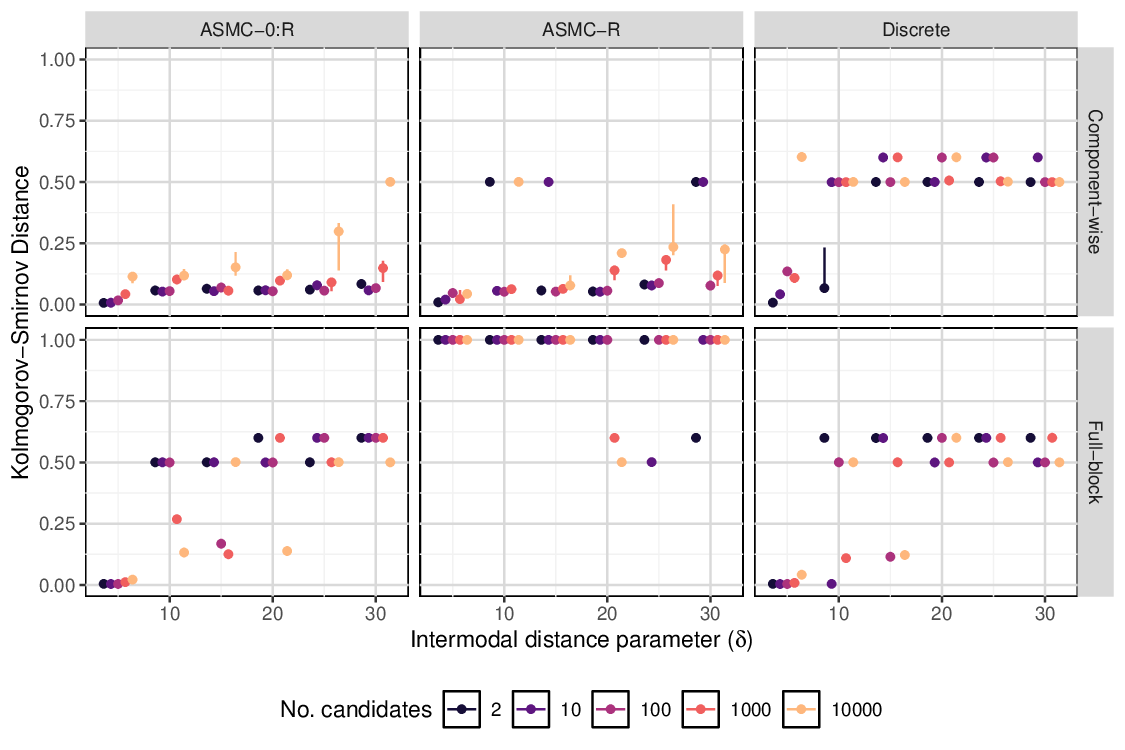}
\caption{Two-sample KSD for the Gaussian mixture distribution across internal CAM configurations. Points represent the median over all independent runs, with thick lines and thin lines denoting the 50\% and 95\% intervals, respectively.}
\label{fig:GM CAM}
\end{figure}

The aggregated KSD results across all six CAM configurations are presented in Figure \ref{fig:GM CAM}. For full-block proposals, performance degraded sharply and inconsistently for $\delta > 5$ across all AGDs, though the discrete and ASMC-0:R AGDs fared slightly better. This suggests that partially uninformed AGDs offer a slight advantage when full-block local proposals struggle. Conversely, component-wise proposals performed much better. Under this configuration, the ASMC-based AGDs consistently outperformed the discrete grid, maintaining stable, low median KSDs even as the intermodal distance increased. While minor KSD inflation and increased cross-run variability emerged at large candidate pools ($M' \ge 1000$), performance remained strong overall. Because the ASMC-R AGD exhibited slightly higher inter-run variance than its pooled counterpart, we select the ASMC-0:R AGD with $M'=10$ as the optimal configuration for external benchmarking.

\begin{figure}[h!]
\centering
\includegraphics[width=\textwidth]{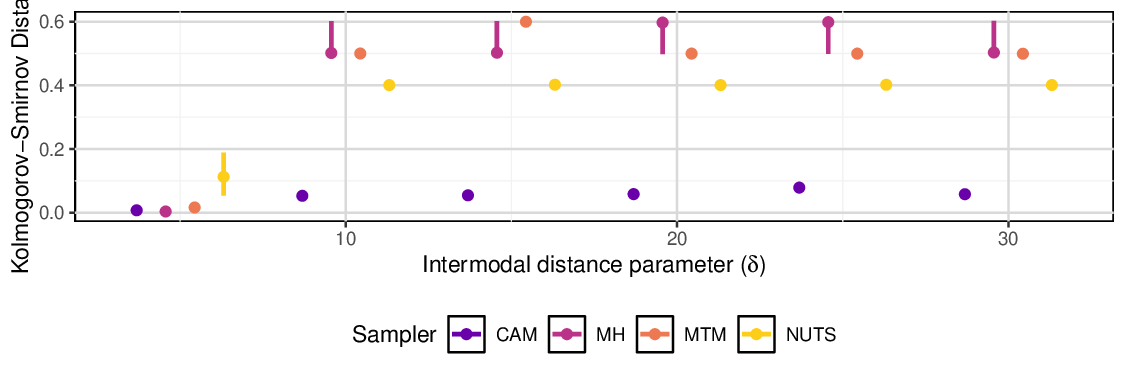}
\caption{Two-sample KSD benchmark comparing CAM against external MCMC methods on the Gaussian mixture distribution. Points represent the median over all runs, with 50\% (thick lines) and 95\% (thin lines) intervals.}
\label{fig:GM KSD}
\end{figure}

The comparison between this CAM configuration and the external MCMC algorithms is shown in Figure \ref{fig:GM KSD}. Because MTM performed uniformly poorly across all candidate sizes, we report its results at $M'=2$ for simplicity. As the intermodal distance grew beyond $\delta=5$, the KSDs for MH, MTM, and Stan approached 1.0, indicating failure to capture the global multimodal structure. In contrast, CAM maintained consistently low KSDs across all tested distances, demonstrating robustness to high degrees of multimodality.

\subsection{Banana Distribution}
\label{sec:banana_distribution}

The banana distribution is a canonical benchmark for evaluating MCMC algorithms due to its highly non-Gaussian curvature \citep{neal2003slice,yang2019componentwise,fontaine2022adaptive}. The target distribution is constructed via a non-linear transformation of a standard normal vector, yielding the unnormalized probability density function
$$
\pi(\mathbf x) \propto \exp\left[-\frac{x_1^2}{200}-\frac{1}{2}\left(x_2+Bx_1^2-100B\right)^2-\frac{1}{2}\sum_{i=3}^d x_i^2\right],
$$
where $\mathbf x = [x_1, \ldots, x_d]^T$ and the parameter $B$ controls the severity of the non-Gaussianity. This parameterization warps the contours of the first two coordinates into a steep, unimodal crescent shape, making them difficult to sample from. We evaluate performance across a gradient of non-Gaussianity, setting $\log_{10}B \in \{-2, -1.75, \ldots, -0.5\}$. Because the adverse geometry is restricted to the first two dimensions, algorithmic efficiency is compared using the ESS per iteration of the first coordinate.

\begin{figure}[h!]
\centering
\includegraphics[width=\textwidth]{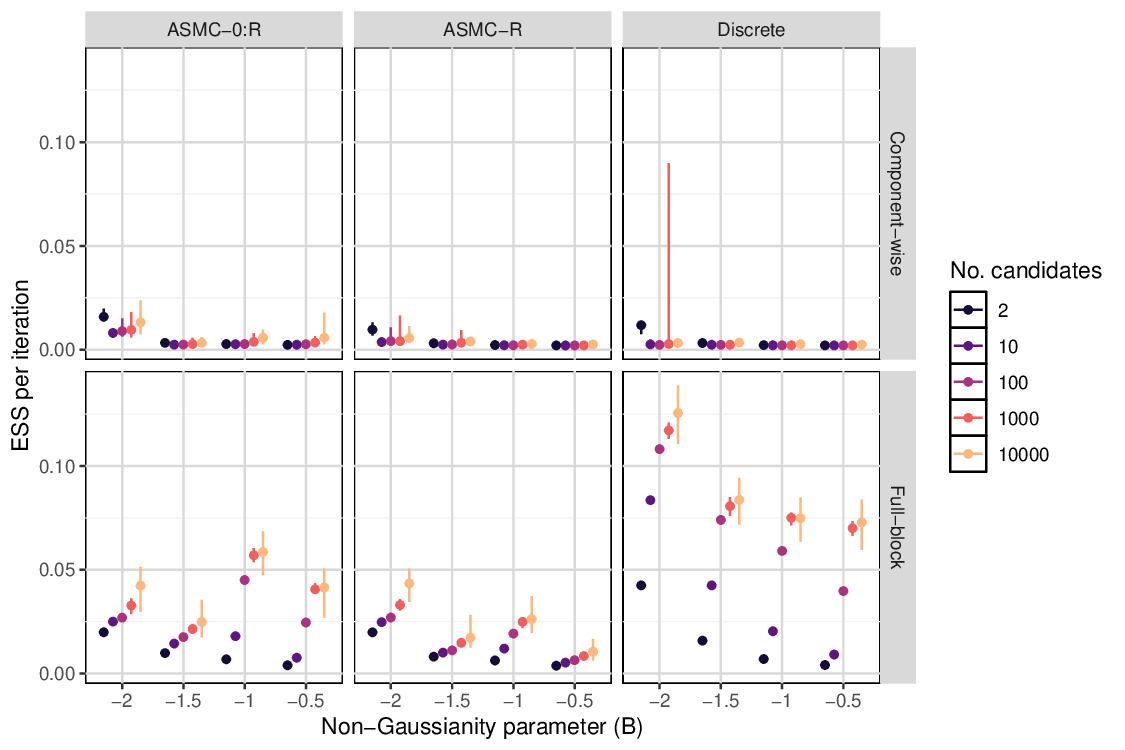}
\caption{ESS per iteration for the first coordinate of the banana distribution across all CAM configurations. Points represent the median over all independent runs, with thick lines and thin lines denoting the 50\% and 90\% intervals, respectively.}
\label{fig:banana CAM}
\end{figure}

Figure \ref{fig:banana CAM} displays the ESS per iteration across all CAM configurations. There is a clear distinction between the proposal architectures: full-block proposals significantly outperform their component-wise counterparts. This result contrasts with a recent review of MTM \citep{doig2025unified} that found that CW proposals are better-suited for non-Gaussian targets. 
Among the full-block configurations, increasing the candidate pool size generally improved sampling efficiency up to $M'=10000$, at which point performance either plateaued or slightly degraded. The discrete uninformed AGD exhibited the most dramatic efficiency gains from larger candidate pools, whereas the informed ASMC-based AGDs yielded more modest marginal benefits. Ultimately, the discrete AGD paired with full-block proposals and $M'=1000$ was selected as the best CAM configuration for external benchmarking, as it achieved near-peak performance without the additional computational overhead of the $M'=10000$ setting.

\begin{figure}[h!]
\centering
\includegraphics[width=\textwidth]{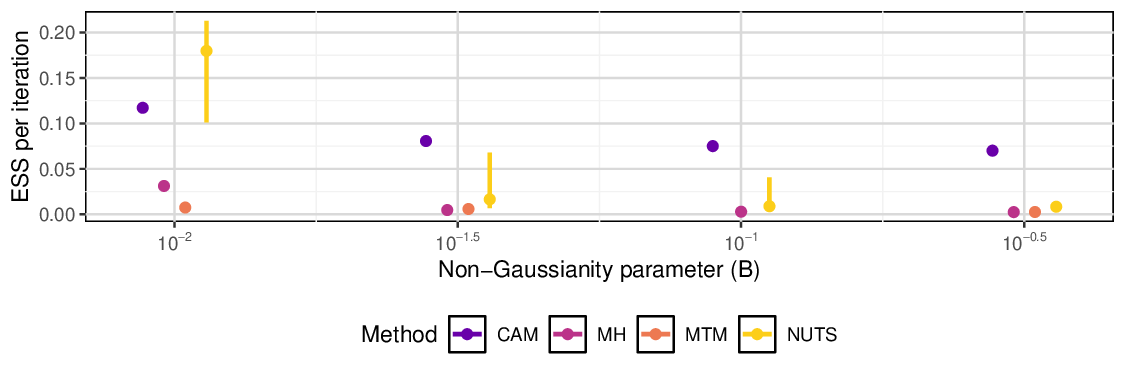}
\caption{ESS per iteration for the first coordinate of the banana distribution, benchmarking CAM against other MCMC methods. Points represent the median over all runs, with 50\% (thick lines) and 90\% (thin lines) intervals.}
\label{fig:banana ESS}
\end{figure}

Figure \ref{fig:banana ESS} benchmarks this CAM configuration against existing methods. For the MTM baseline, $M=10000$ yielded the best relative performance. Both MH and MTM exhibited particularly poor efficiency compared to CAM and Stan. MTM failed to converge ($\hat{R} > 1.05$) for the severe $B=10^{-1}$ setting, resulting in no valid data for that coordinate. These failures reinforce the observation that standard component-wise random walk struggles in highly non-Gaussian spaces unless assisted by global auxiliary jumps. While NUTS achieved the highest ESS per iteration for milder geometries (lower $B$), its performance degraded sharply as the target curvature became more extreme. CAM however, maintained consistent sampling efficiency across the entire non-Gaussianity gradient, surpassing the performance of NUTS for the more severely curved target distributions.

\subsection{Eight Schools Model}
\label{sec:eight_schools}

To evaluate performance on a Bayesian inference problem derived from real data, we use the eight schools problem, a canonical Bayesian hierarchical model. This is a model for a meta-analysis of the estimated effects of SAT coaching programs across eight different high schools \citep{rubin1981eightschools}. For each school $i = 1, \dots, 8$, an observed treatment effect $y_i$ and its associated standard error $\sigma_i$ are provided. The effects are assumed to follow a normal distribution with a latent, school-specific mean $\theta_i$, which in turn has a shared population-level hyperprior distribution governed by a global mean $\mu$ and global standard deviation $\tau$. The target posterior distribution is defined by the following hierarchy \citep{gelman2011BDA}:
\begin{align*}
    \mu &\sim \mathcal N(0,5^2), & \tau &\sim \text{Cauchy}_+(0,5), \\
    \theta_i &\sim \mathcal N(\mu,\tau^2), & y_i &\sim \mathcal N(\theta_i,\sigma_i^2).
\end{align*}
After discarding runs that failed to converge (i.e., retaining only those with an average $\hat R < 1.05$), we computed the ESS per iteration for each of the ten model parameters.

\begin{figure}[h!]
\centering
\includegraphics[width=\textwidth]{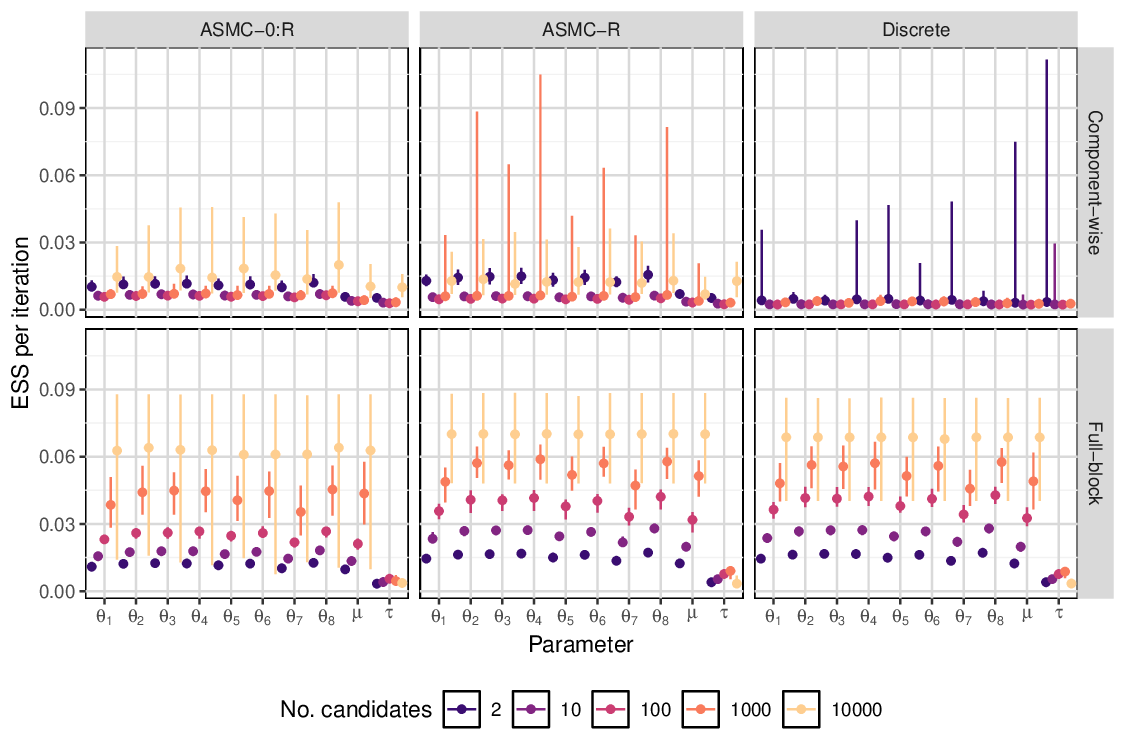}
\caption{ESS per iteration for each parameter in the eight schools model across internal CAM configurations. Points represent the median over converged runs, with thick and thin vertical lines denoting the 50\% and 90\% intervals, respectively.}
\label{fig:ES CAM}
\end{figure}

The ESS per iteration for the internal CAM configurations is presented in Figure \ref{fig:ES CAM}. Across all parameters, component-wise proposals performed uniformly worse than full-block proposals. Within the full-block architectures, sampling efficiency generally increased as the number of candidates $M'$ increased, demonstrating relatively consistent performance across the different AGDs. However, the component-wise architectures exhibited a non-linear relationship with candidate size: initial increases in $M'$ degraded per-iteration efficiency, with $M'=10000$ being the only pool size to eventually surpass the $M'=2$ baseline. While sampling efficiency was highly consistent across the location parameters ($\mu$ and the $\theta_i$'s), the global scale parameter $\tau$ presented a distinct challenge due to the funnel-like geometry it induces in the hierarchical posterior. Although component-wise proposals showed little discrepancy in efficiency between the mean parameters and $\tau$, the full-block proposals yielded a markedly lower ESS per iteration for $\tau$. Notably, as $M'$ grew to 10000, the full-block proposals suffered a sharp decrease in median efficiency for this variance parameter. Balancing these dynamics, we selected the CAM configuration using the ASMC-0:R AGD with $M=10000$ candidates for subsequent benchmarking.

\begin{figure}[h!]
\centering
\includegraphics[width=\textwidth]{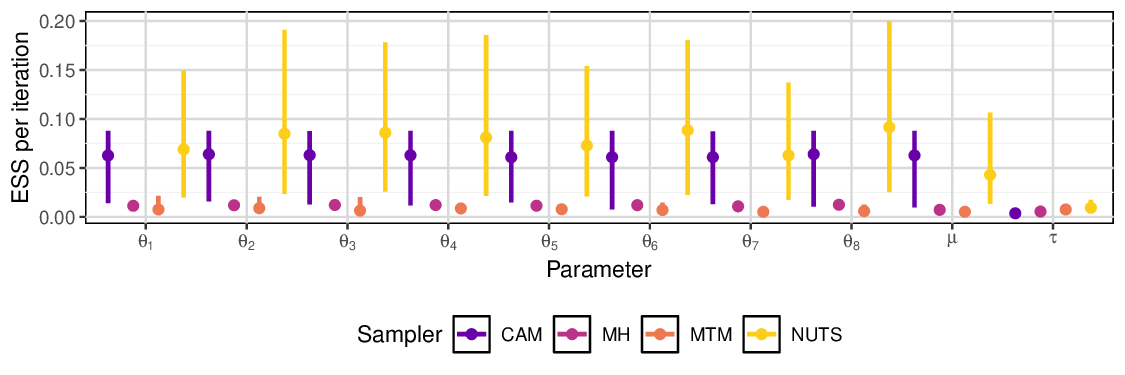}
\caption{Benchmark ESS per iteration for each parameter in the eight schools model. Points represent the median across 50 independent runs, with thick and thin vertical lines denoting the 50\% and 90\% intervals, respectively.}
\label{fig:ES ESS}
\end{figure}

The comparison between this CAM configuration and other MCMC methods is shown in Figure \ref{fig:ES ESS}. For the location parameters ($\mu, \theta_i$), CAM and Stan demonstrated comparable per-iteration sampling efficiency, though Stan occasionally obtained a higher upper-bound ESS. With respect to the challenging variance parameter $\tau$, Stan and standard MTM achieved higher per-iteration efficiency. Overall, these results are encouraging: while NUTS (Stan) is optimized for the continuous, unimodal geometries typical of hierarchical models, CAM remains competitive. Taken together with previous experiments, this demonstrates that CAM sacrifices little to no efficiency on standard inference tasks while offering vastly superior robustness on pathological, multimodal, and severely non-Gaussian distributions.

\section{Discussion}
\label{sec:discussion P2}

In this paper, we introduced Compound Auxiliary Metropolis, a general MCMC framework that explicitly incorporates auxiliary information into a multi-candidate proposal mechanism. By including auxiliary variables through an auxiliary generating distribution, CAM allows users to tailor the transition kernel to the challenges presented by their specific target distributions. This algorithm is a straightforward extension of the multiple-try Metropolis framework, opening it up to a broader class of auxiliary-enhanced MCMC algorithms. Furthermore, we frame CAM as an instance of involutive MCMC, demonstrating that CAM is a valid MCMC algorithm that preserves the detailed balance condition.

The efficiency gains enabled by CAM were empirically validated through simulation. We chose target distributions that represent known challenges for modern MCMC methods: 1) surfaces exhibiting sharply non-Gaussian curvature, even at very local scales; and 2) distributions containing multiple well-separated modes. In all experimental settings, the integration of auxiliary information into the multi-candidate framework demonstrated consistent performance gains over standard MTM. Moreover, while CAM and the No-U-Turn Sampler (NUTS, via Stan) exhibited comparable efficiency on the relatively mild challenges presented by the funnel-shaped posterior of the eight schools hierarchical model and some of the simulation settings, CAM outperformed NUTS as the target topographies became more extreme.

Our empirical results provide guidance for configuring CAM based on the features of the target distribution. For multimodal targets, as in our Gaussian mixture experiment, it is best to use component-wise state updates paired with an informed auxiliary variable to explore multiple, widely separated modes. However, there is a trade-off: while informed auxiliaries (like those derived from ASMC) can improve sampling efficiency, their benefit is limited by the quality of the initial approximation. If the informed AGD drops modes, the chain may fail to discover them. In such cases, using an uninformed discrete grid, with well-chosen boundaries, ensures a more complete exploration of the state-space, albeit at a lower per-iteration efficiency. When the target exhibits sharply non-Gaussian curvature at local scales (such as the banana distribution), a different configuration is optimal. Here, full-block updates outperform component-wise updates, regardless of the auxiliary chosen. Among the auxiliaries used in the study, uninformed auxiliaries were the most efficient. Uninformed AGDs propose transitions that are entirely independent of the chain's current position, thus producing a chain with relatively low autocorrelation and therefore a higher effective sample size per iteration. As the severe geometry of the banana distribution often produces highly auto-correlated MCMC chains, the benefit of using an uninformed auxiliary is emphasized in this setting.

CAM an open framework for future algorithm development. The specific AGDs presented in this paper, a discrete grid and empirical distributions from ASMC, are proofs-of-concept. The biggest limitation of the current framework is that the possible benefits from using CAM are limited by the quality of the auxiliary variables. When the auxiliary information is sparse or poorly constructed, CAM's performance is on par with standard MTM. Using more effective methods to construct the auxiliary distribution, such as variational inference or normalizing flows, could further improve CAM's performance. Furthermore, the auxiliary generating distribution could in principle be improved dynamically. These are two examples of how CAM, as it has been presented here, may be extended to a broader and more effective class of MCMC methods.

\bigskip
\begin{center}
{\large\bf SUPPLEMENTARY MATERIAL}
\end{center}

\section*{Supplementary A~~~~Annealed Sequential Monte Carlo}
\label{sec:ASMC P0}

\renewcommand{\theequation}{A\arabic{equation}}
\renewcommand{\thealgorithm}{A\arabic{algorithm}}

In this appendix, we provide a detailed description of the annealed sequential Monte Carlo algorithm \citep{delmoral2006SMCsampler} followed by the corresponding pseudocode. ASMC is an advanced importance sampling method designed to sample from a complex target distribution $\pi$, defined on a state space $\mathcal X$, by propagating particles through a sequence of tempered intermediate distributions. 

Typically, the target distribution is known only up to an intractable normalizing constant $Z$. Let $\gamma$ denote the unnormalized density of $\pi$, such that $\pi(\mathbf x) = \gamma(\mathbf x)/Z$ for $\mathbf x \in \mathcal X$. The tempered intermediate distributions take the geometric form:
\begin{equation*}
    \pi_\alpha(\mathbf x) \propto \gamma_\alpha(\mathbf x) = \left[\gamma(\mathbf x)\right]^\alpha\left[\rho(\mathbf x)\right]^{1-\alpha},
\end{equation*}
where $\alpha \in [0,1]$ is an annealing parameter and $\rho$ is a tractable reference distribution from which independent samples can be readily drawn. This formulation constructs a continuous geometric path connecting the simple reference distribution ($\alpha=0$) to the complex target distribution ($\alpha=1$). In practice, the ASMC algorithm discretizes this path using an increasing, finite sequence of $R+1$ annealing parameters, $0=\alpha_0 < \alpha_1 < \dots < \alpha_R=1$, producing the specific sequence of intermediate distributions:
\begin{equation}\label{eqn:intermediate distribution P0}
    \pi_r(\mathbf x) \propto \gamma_r(\mathbf x) = \left[\gamma(\mathbf x)\right]^{\alpha_r}\left[\rho(\mathbf x)\right]^{1-\alpha_r}, \quad r=0,\ldots,R.
\end{equation}
Note that the sequence begins at the reference distribution ($\pi_0 = \rho$) and terminates at the target distribution ($\pi_R = \pi$).

The algorithm is initialized by drawing $K$ independent particles from the reference distribution $\rho$. Because these initial samples are independent samples, all $K$ particles are assigned a uniform normalized importance weight of $1/K$:
$$\left\{\mathbf x_{0,k},\frac{1}{K}\right\}_{k=1}^K.$$
To generate samples from subsequent distributions $\pi_r$ ($r \ge 1$), this weighted particle ensemble is iteratively updated via adaptive weight computations, resampling when necessary, and Markov transitions. For a specific transition from $\pi_{r-1}$ to $\pi_r$, the algorithm first computes the \textit{incremental importance weight} for each particle based on its current position $\mathbf x_{r-1,k}$:
\begin{equation}\label{eqn:incremental_weight}
    \tilde w_{r,k} = \frac{\gamma_r(\mathbf x_{r-1,k})}{\gamma_{r-1}(\mathbf x_{r-1,k})}.
\end{equation}
The unnormalized cumulative weight is then updated recursively via $w_{r,k} = \tilde w_{r,k} \cdot w_{r-1,k}$, and the normalized weights $W_{r,k}$ are computed such that $\sum_{k=1}^K W_{r,k} = 1$. Because the incremental weight updates depend only on the particles' previous positions, the annealing step $\Delta\alpha_r = \alpha_r - \alpha_{r-1}$ can be determined adaptively at the beginning of the $r$-th iteration.

Treating the complete sequence $\{\alpha_r\}_{r=0}^R$ as a tuning parameter is both difficult and time-consuming. Instead, ASMC uses an adaptive procedure based on the conditional effective sample size, which quantifies how well the weighted sample from $\pi_{r-1}$ estimates expectations under the subsequent distribution $\pi_r$ \citep{zhou2016toward}. Specifically, the algorithm monitors the relative conditional effective population size (rCESS), scaled to $[1/K, 1]$:
\begin{equation}\label{eqn:rCESS P0}
    \text{rCESS}_r\left(W_{r-1,1:K},\tilde w_{r,1:K}\right) = \frac{\left(\sum_{k=1}^K W_{r-1,k}\tilde w_{r,k}\right)^2}{\sum_{k=1}^K W_{r-1,k}\tilde w_{r,k}^2}.
\end{equation}
Because the incremental weights $\tilde w_{r,k}$ are a function of only $\mathbf x_{r-1,1:K}$ and the proposed step size $\Delta\alpha_r$, we can enforce a constraint on efficiency by setting a desired rCESS threshold, $\phi \in (0,1)$. A standard bisection algorithm is used to find the $\Delta\alpha_r$ that satisfies:
\begin{equation}\label{eqn:bisection P0}
    f(\Delta\alpha) = \text{rCESS}_r\left(W_{r-1,1:K}, \left\{ \left[ \frac{\gamma_{r}(\mathbf x_{r-1,k})}{\gamma_{r-1}(\mathbf x_{r-1,k})} \right]^{\Delta\alpha} \right\}_{k=1}^K \right) = \phi.
\end{equation}
This adaptation reduces the tuning burden from choosing $R$ separate parameters to selecting a single scalar $\phi$. Values of $\phi > 0.9$ yield small steps and very long, stable annealing sequences, while lower values produce shorter, faster sequences at the risk of higher estimator variance. For severe pathological targets, values very close to 1 are often required \citep{wang2020ASMCphylo}. In our experiments we use a value of $\phi=0.8$ to quickly generate a coarse approximation to the target.

Following the weight update, the algorithm must manage weight degeneracy--a phenomenon where the cumulative importance weight inevitably concentrates on a vanishingly small fraction of the particles \citep{doucet2009tutorial}. To quantify this, the relative effective sample size (rESS) of the current normalized weights is evaluated:
\begin{equation}\label{eqn:rESS P0}
\text{rESS}_r = \left[K\sum_{k=1}^K\left(W_{r,k}\right)^2\right]^{-1}.
\end{equation}
When the rESS drops below a predefined threshold $\epsilon$ (typically $\epsilon = 0.5$), a resampling step is triggered. Particles are resampled with replacement probability proportional to their weights $W_{r,k}$, effectively duplicating high-weight particles and discarding low-weight ones. Following resampling, all weights are reset to uniformity ($w_{r,k} = 1$, $W_{r,k} = 1/K$). While resampling mitigates weight degeneracy, it introduces particle degeneracy (loss of unique particle diversity) and increases the variance of the estimators \citep{chopin2004central}. To minimize this induced variance, we use the systematic resampling algorithm \citep{douc2005comparison}.

Finally, to restore particle diversity, each particle is independently moved by applying an $\pi_r$-invariant transition kernel: $\mathbf x_{r,k} \sim T_r(\cdot|\mathbf x_{r-1,k})$. In this work, $T_r$ is a standard reversible MCMC kernel, though other mechanisms are possible \citep{delmoral2006SMCsampler}. The complete ASMC procedure is detailed in Algorithm \ref{algo:ASMC}.

\begin{algorithm}[h!]
   \caption{\bf{Annealed Sequential Monte Carlo}}
  \label{algo:ASMC}
{
\begin{algorithmic}[1]
\State {\bfseries Input:} (a) Target unnormalized density $\gamma$, (b) Target rCESS $\phi$, (c) Number of particles $K$.
\State {\bfseries Settings:} Resampling threshold $\epsilon=0.5$.
\State {\bfseries Output:} Annealing schedule $\{\alpha_r\}_{r=0}^R$ and weighted particles $\left\{\mathbf x_{r,k},W_{r,k}\right\}_{r=0,k=1}^{R,K}$.
\For{$k=1,\ldots,K$}
    \State Initialize particles $\mathbf x_{0,k} \sim \rho$ and weights $w_{0,k}=1$, $W_{0,k}=1/K$.
\EndFor
\State Initialize annealing schedule $\alpha_0 \gets 0$ and step counter $r \gets 0$.
\While{$\alpha_r < 1$}
    \State $r \gets r+1$
    \State Obtain $\Delta\alpha_r$ by solving Equation \ref{eqn:bisection P0} such that $\alpha_r = \min(1, \alpha_{r-1} + \Delta\alpha_r)$.
    \State Update unnormalized weights $w_{r,k}$ via Equation \ref{eqn:incremental_weight}.
    \State Compute normalized weights $W_{r,1:K}$.
    \State Compute $\text{rESS}_r(W_{r,1:K})$ via Equation \ref{eqn:rESS P0}.
    \If{$\text{rESS}_r < \epsilon$}
        \State Resample indices from $\{1, \dots, K\}$ with probabilities $W_{r,1:K}$ (via systematic resampling).
        \State Update particles $\mathbf x_{r,1:K}$ based on resampled indices.
        \State Reset weights: $w_{r,k} \gets 1$ and $W_{r,k} \gets 1/K$.
    \EndIf
    \For{$k=1,\ldots,K$}
        \State Apply transition kernel to update particle: $\mathbf x_{r,k} \sim T_r(\cdot|\mathbf x_{r-1,k})$.
    \EndFor
\EndWhile
\State \Return $\{\alpha_r\}_{r=0}^R$ and $\left\{\mathbf x_{r,k},W_{r,k}\right\}_{r=0,k=1}^{R,K}$.
\end{algorithmic}
}
\end{algorithm}

\section*{Supplementary B~~~~Multiple-Try Metropolis}

\renewcommand{\theequation}{B\arabic{equation}}
\renewcommand{\thealgorithm}{B\arabic{algorithm}}

In this section of the Supplementary Materials we describe the multiple-try Metropolis algorithm \citep{liu2000mtm}. Suppose a Markov chain targeting $\pi$ is currently at state $\mathbf x$. The compound proposal characterizing MTM comprises $M$ candidates drawn from individual proposal distributions: $\mathbf y_m \sim T_m(\cdot|\mathbf x)$ for $m=1,\ldots,M$. A single candidate $\mathbf y_J$ is then selected from this pool with probability $P(J|\mathbf y_{1:M},\mathbf x)$. The conventional design of MTM algorithms assigns a weight $u_m(\mathbf y_m,\mathbf x)$ to each candidate, defining the selection probability as:
\begin{equation}\label{eqn:MTM selection probability}
    P(J|\mathbf y_{1:M},\mathbf x) = \frac{u_J(\mathbf y_J,\mathbf x)}{\sum_{m=1}^M u_m(\mathbf y_m,\mathbf x)}.
\end{equation}
The functional form of these weights is well-represented in the MTM literature \citep{liu2000mtm,gagnon2023localbalancing,pandolfi2014genMTMpap,yang2019componentwise}.

To satisfy detailed balance and construct a reversible MCMC kernel, a ``reverse'' candidate selection probability must be computed. This requires specifying $M$ reverse samples, denoted $\mathbf x_m^*$. The $J$-th reverse sample is set to the current state, $\mathbf x_J^* = \mathbf x$, while the remaining $M-1$ reverse samples are drawn conditioned on the selected proposal: $\mathbf x_m^* \sim T_m(\cdot|\mathbf y_J)$ for $m \ne J$. Finally, the selected proposal $\mathbf y_J$ is accepted as the next state with probability:
\begin{equation}\label{eqn:MTM acceptance probability}
    a(\mathbf x,\mathbf y_J) = 1 \land \frac{\pi(\mathbf y_J)T_J(\mathbf x|\mathbf y_J)P(J|\mathbf x_{1:M}^*,\mathbf y_J)}{\pi(\mathbf x)T_J(\mathbf y_J|\mathbf x)P(J|\mathbf y_{1:M},\mathbf x)}.
\end{equation}
If rejected, the chain remains at state $\mathbf x$. The pseudocode for MTM is given in Algorithm \ref{algo:MTM} below.
\begin{algorithm}[h!]
\caption{\bf{Multiple-Try Metropolis}}
\label{algo:MTM}
\begin{algorithmic}[1]
\State {\bfseries Input:} (a) Target distribution $\pi$; (b) current state $\mathbf x$. 
\State {\bfseries Settings:} (a) Number of candidates $M$; (b) proposal distributions $\{T_m\}$; (c) weight functions $\{u_m\}$.
\For{$m \in \{1,\ldots,M\}$}
    \State Generate candidate $\mathbf y_m \sim T_m(\cdot|\mathbf x)$.
    \State Compute forward weights $u_m(\mathbf y_m, \mathbf x)$.
\EndFor
\State Sample $J \in \{1,\ldots,M\}$ with probability given by Equation \ref{eqn:MTM selection probability}.
\For{$m \in \{1, 2, \dots, M\}$}
    \If{$m=J$}
        \State Set $\mathbf x_m^* = \mathbf x$.
    \Else
        \State Generate reverse samples $\mathbf x_m^* \sim T_m(\cdot|\mathbf y_J)$.
    \EndIf
    \State Compute reverse weights $u_m(\mathbf x_m^*, \mathbf y_J)$.
\EndFor
\State Compute the acceptance probability $a(\mathbf x, \mathbf y_J)$ using Equation \ref{eqn:MTM acceptance probability}.
\State Sample $U \sim \text{Unif}(0,1)$.
\If{$U < a(\mathbf x, \mathbf y_J)$}
    \State Return $\mathbf y_J$.
\Else
    \State Return $\mathbf x$.
\EndIf
\end{algorithmic}
\end{algorithm}

\section*{Supplementary C~~~~Balanced Selection Adaptation}
\label{sec:balanced selection}

\renewcommand{\theequation}{C\arabic{equation}}
\renewcommand{\thealgorithm}{C\arabic{algorithm}}

This appendix reviews the balanced selection rate adaptation scheme designed for component-wise multiple-try Metropolis \citep{yang2019componentwise}. This adaptive procedure tunes the local proposal variances to ensure that candidates drawn from a sequence of variance scales are selected with well-balanced frequencies, preventing the chain from collapsing into overly narrow or excessively diffuse exploration.

Let $\mathbf x \in \mathbb{R}^d$ denote the current state, with $x_i$ representing its $i$-th coordinate. During a component-wise update, local candidates for this coordinate are drawn independently from univariate Gaussian distributions: $y_{i,m} \sim \mathcal N(x_i, \sigma_{i,m}^2)$, where $\sigma_{i,m}$ is the standard deviation of the $m$-th proposal. For any given coordinate $i$, the candidate standard deviations $\{\sigma_{i,m}\}_{m=1}^M$ are constrained to remain equally spaced on the $\log_2$ scale. 

Adaptation of the proposal variance can reduce long-run effective sample sizes by introducing autocorrelation or potentially risking the ergodicity of the chain \citep{andrieu2008adaptiveMCMC}. To mitigate this, adaptation occurs only every $\beta$ iterations, subject to a diminishing adaptation probability
\begin{equation}\label{eqn:adaptation prob}
    P(n) = \max\left(0.99^{a(n)-1}, a(n)^{-1/2}\right),
\end{equation}
where $a(n) = (n-\beta)/\beta$ and $n$ is the current iteration. 
Between these adaptation events, the empirical selection rate $S_{i,m}$ for each candidate $m$ at coordinate $i=1,\ldots,d$ is recorded. When adaptation is triggered, the extreme variances (the highest, $\sigma_{i,M}$, and the lowest, $\sigma_{i,1}$) are adjusted based on their selection rates relative to a uniform selection rate of $1/M$. 
If $S_{i,M} > 2/M$, the largest variance is too frequently selected, so $\sigma_{i,M}$ is doubled. If $S_{i,M} < 1/(2M)$, then it is selected too rarely, so $\sigma_{i,M}$ is halved. On the other hand, if $S_{i,1} > 2/M$, the smallest variance is selected too often, so $\sigma_{i,1}$ is halved. If $S_{i,1} < 1/(2M)$, it is doubled. 
If either $\sigma_{i,1}$ or $\sigma_{i,M}$ is updated, the remaining standard deviations $\sigma_{i,1:M}$ are re-interpolated to maintain equal spacing on the $\log_2$ scale. To ensure numerical stability, the standard deviations are bounded within $[2^\epsilon, 2^L]$. Following the recommendations of the original authors \citep{yang2019componentwise}, we use hyperparameter values of $\beta=100$, $\epsilon=-15$, and $L=50$. Further discussion of this is available in the original manuscript, and a comparative analysis of its performance within MTM frameworks can be found in \citep{doig2025unified}. The complete procedure is detailed in Algorithm \ref{algo:balanced selection}.

\begin{algorithm}[h!]
   \caption{\bf{Balanced Selection Component-Wise Adaptation}}
  \label{algo:balanced selection}
{
\begin{algorithmic}[1]
\State {\bfseries Input:} (a) Current state $\mathbf x$, (b) current proposal standard deviations $\{\sigma_{i,m}\}$, and (c) current iteration $n$.
\State {\bfseries Settings:} $\beta=100$, $\epsilon=-15$, and $L=50$.
\State {\bfseries Output:} Updated proposal standard deviations, $\{\sigma_{i,m}\}$.
\State Compute the adaptation probability $P(n)$ (Equation \ref{eqn:adaptation prob}).
\State Sample $U \sim \text{Uniform}(0,1)$.
\If{$(n \mod \beta == 0)$ \textbf{and} $(U < P(n))$}
     \For{$i=1,\ldots,d$}
        \If{$S_{i,M} > 2/M$}
            \State $\sigma_{i,M} = \min(2\sigma_{i,M}, 2^L)$.
        \ElsIf{$(S_{i,M} < 1/(2M))$ \textbf{and} $(\sigma_{i,M}/2 > \sigma_{i,1})$}
            \State $\sigma_{i,M} = \max(\sigma_{i,M}/2, 2^\epsilon)$.
        \EndIf
        
        \If{$S_{i,1} > 2/M$}
            \State $\sigma_{i,1} = \max(\sigma_{i,1}/2, 2^\epsilon)$.
        \ElsIf{$(S_{i,1} < 1/(2M))$ \textbf{and} $(2\sigma_{i,1} < \sigma_{i,M})$}
            \State $\sigma_{i,1} = \min(2\sigma_{i,1}, 2^L)$.
        \EndIf
        
        \State Rescale $\sigma_{i,1:M}$ to be equally spaced on the $\log_2$ scale.
        \State Reset selection rates $S_{i,1:M} = 0$.
     \EndFor
\EndIf
\State \Return $\{\sigma_{i,m}\}_{i=1:d, m=1:M}$.
\end{algorithmic}
}
\end{algorithm}






\bibliographystyle{chicago}
\bibliography{references}

\begin{thebibliography}{}

\bibitem[\protect\citeauthoryear{Andrieu and Thoms}{Andrieu and Thoms}{2008}]{andrieu2008adaptiveMCMC}
Andrieu, C. and J.~Thoms (2008).
\newblock A tutorial on adaptive {MCMC}.
\newblock {\em Statistical Computing\/}~{\em 18}, 343--373.

\bibitem[\protect\citeauthoryear{Bezanson, Edelman, Karpinski, and Shah}{Bezanson et~al.}{2017}]{julia}
Bezanson, J., A.~Edelman, S.~Karpinski, and V.~B. Shah (2017).
\newblock Julia: {A} fresh approach to numerical computing.
\newblock {\em SIAM review\/}~{\em 59\/}(1), 65--98.

\bibitem[\protect\citeauthoryear{Biron-Lattes, Surjanovic, Syed, Campbell, and Bouchard-C\^ot\'e}{Biron-Lattes et~al.}{2024}]{bironlattes2024automala}
Biron-Lattes, M., N.~Surjanovic, S.~Syed, T.~Campbell, and A.~Bouchard-C\^ot\'e (2024, 02--04 May).
\newblock {autoMALA}: Locally adaptive {M}etropolis-adjusted {L}angevin algorithm.
\newblock In S.~Dasgupta, S.~Mandt, and Y.~Li (Eds.), {\em Proceedings of The 27th International Conference on Artificial Intelligence and Statistics}, Volume 238 of {\em Proceedings of Machine Learning Research}, Valencia, Spain, pp.\  4600--4608. PMLR.

\bibitem[\protect\citeauthoryear{Casarin, Craiu, and Leisen}{Casarin et~al.}{2013}]{casarin2013interactingMTM}
Casarin, R., R.~Craiu, and F.~Leisen (2013).
\newblock Interacting multiple try algorithms with different proposal distribution.
\newblock {\em Statistical Computing\/}~{\em 23}, 185--200.

\bibitem[\protect\citeauthoryear{Chopin}{Chopin}{2004}]{chopin2004central}
Chopin, N. (2004).
\newblock Central limit theorem for sequential {M}onte {C}arlo methods and its application to {B}ayesian inference.
\newblock {\em The Annals of Statistics\/}~{\em 32\/}(6), 2385--2411.

\bibitem[\protect\citeauthoryear{Doig and Wang}{Doig and Wang}{2026}]{doig2025unified}
Doig, R. and L.~Wang (2026).
\newblock A unified framework for multiple-try {M}etropolis: Construction and empirical benchmarks.

\bibitem[\protect\citeauthoryear{Douc and Capp{\'e}}{Douc and Capp{\'e}}{2005}]{douc2005comparison}
Douc, R. and O.~Capp{\'e} (2005).
\newblock Comparison of resampling schemes for particle filtering.
\newblock In {\em Image and Signal Processing and Analysis, 2005. ISPA 2005. Proceedings of the 4th International Symposium on}, pp.\  64--69. IEEE.

\bibitem[\protect\citeauthoryear{Doucet and Johansen}{Doucet and Johansen}{2009}]{doucet2009tutorial}
Doucet, A. and A.~Johansen (2009).
\newblock A tutorial on particle filtering and smoothing: Fifteen years later.
\newblock {\em Handbook of Nonlinear Filtering\/}~{\em 12}.

\bibitem[\protect\citeauthoryear{Edwards and Sokal}{Edwards and Sokal}{1988}]{edwards1988FKSWIsing}
Edwards, R.~G. and A.~D. Sokal (1988).
\newblock Generalization of the {F}ortuin-{K}asteleyn-{S}wendsen-{W}ang representation and {M}onte {C}arlo algorithm.
\newblock {\em Physical Review Letters\/}~{\em 38}, 2009--2012.

\bibitem[\protect\citeauthoryear{Fontaine and B\'{e}ard}{Fontaine and B\'{e}ard}{2022}]{fontaine2022adaptive}
Fontaine, S. and M.~B\'{e}ard (2022).
\newblock An adaptive multiple-try {M}etropolis algorithm.
\newblock {\em Bernoulli\/}~{\em 28\/}(3), 1986--2011.

\bibitem[\protect\citeauthoryear{Gagnon, Maire, and Zanella}{Gagnon et~al.}{2023}]{gagnon2023localbalancing}
Gagnon, P., F.~Maire, and G.~Zanella (2023).
\newblock Improving multiple-try {M}etropolis with local balancing.
\newblock {\em Journal of Machine Learning Research\/}~{\em 24}, 1--59.

\bibitem[\protect\citeauthoryear{Gelman, Carlin, Stern, Dunson, Vehtari, and Rubin}{Gelman et~al.}{2013}]{gelman2011BDA}
Gelman, A., J.~Carlin, H.~Stern, D.~Dunson, A.~Vehtari, and D.~Rubin (2013).
\newblock {\em Bayesian Data Analysis\/} (3 ed.).
\newblock Boca Raton, USA: Chapman Hall/CRC.

\bibitem[\protect\citeauthoryear{Goedman and {others}}{Goedman and {others}}{2024}]{goedman2012stanjulia}
Goedman, R.~J. and {others} (2024).
\newblock Stan.jl.

\bibitem[\protect\citeauthoryear{Goodman and Weare}{Goodman and Weare}{2010}]{goodman2010ensemble}
Goodman, J. and J.~Weare (2010).
\newblock Ensemble samplers with affine invariance.
\newblock {\em Communications in Applied Mathematics and Computational Science\/}~{\em 5\/}(1), 65--80.

\bibitem[\protect\citeauthoryear{Green}{Green}{1992}]{green1992SW}
Green, P.~J. (1992).
\newblock A note on the {S}wendsen-{W}ang algorithm for ordered colours.
\newblock Technical report, University of Bristol, Statistics Group.

\bibitem[\protect\citeauthoryear{Gu, Ghahramani, and Turner}{Gu et~al.}{2015}]{gu2015neuralSMC}
Gu, S., Z.~Ghahramani, and R.~Turner (2015).
\newblock Neural adaptive sequential {M}onte {C}arlo.
\newblock In {\em Advances in Neural Information Processing Systems}.

\bibitem[\protect\citeauthoryear{Hastings}{Hastings}{1970}]{hastings1970}
Hastings, W.~K. (1970).
\newblock Monte {C}arlo sampling methods using {M}arkov chains and their applications.
\newblock {\em Biometrika\/}~{\em 57\/}(1), 97--109.

\bibitem[\protect\citeauthoryear{Higdon}{Higdon}{1998}]{higdon1998auxvarbinary}
Higdon, D.~M. (1998).
\newblock Auxiliary variable methods for {M}arkov chain {M}onte {C}arlo with applications.
\newblock {\em Journal of the American Statistical Association\/}~{\em 93\/}(442).

\bibitem[\protect\citeauthoryear{Hoffman and Gelman}{Hoffman and Gelman}{2014}]{hoffman2014nuts}
Hoffman, M.~D. and A.~Gelman (2014).
\newblock The {N}o-{U}-{T}urn {S}ampler: Adaptively setting path lengths in {H}amiltonian {M}onte {C}arlo.
\newblock {\em Journal of Machine Learning Research\/}~{\em 15}, 1593--1623.

\bibitem[\protect\citeauthoryear{Liang and Wong}{Liang and Wong}{2001}]{liang2001real}
Liang, F. and W.~H. Wong (2001).
\newblock Real-parameter evolutionary {M}onte {C}arlo with applications to {B}ayesian mixture models.
\newblock {\em Journal of the American Statistical Association\/}~{\em 96\/}(454), 653--666.

\bibitem[\protect\citeauthoryear{Liu, Liang, and Wong}{Liu et~al.}{2000}]{liu2000mtm}
Liu, J.~S., F.~Liang, and W.~H. Wong (2000).
\newblock The multiple-try method and local optimization in {M}etropolis sampling.
\newblock {\em Journal of the American Statistical Association\/}~{\em 95\/}(449), 121--134.

\bibitem[\protect\citeauthoryear{Marinari and Parisi}{Marinari and Parisi}{1992}]{marinari1992simulatedtempering}
Marinari, E. and G.~Parisi (1992).
\newblock Simulated tempering: a new {M}onte {C}arlo scheme.

\bibitem[\protect\citeauthoryear{Martino}{Martino}{2018}]{martino2018review}
Martino, L. (2018).
\newblock A review of multiple try {MCMC} algorithms for signal processing.
\newblock {\em Digital Signal Processing\/}~{\em 75}, 134--152.

\bibitem[\protect\citeauthoryear{Metropolis, Rosenbluth, Rosenbluth, Teller, and Teller}{Metropolis et~al.}{1953}]{metropolis1953}
Metropolis, N., A.~W. Rosenbluth, M.~Rosenbluth, A.~H. Teller, and E.~Teller (1953).
\newblock Equation of state calculations by fast computing machines.
\newblock {\em Journal of Chemical Physics\/}~{\em 21}, 1087--1092.

\bibitem[\protect\citeauthoryear{Moral, Doucet, and Jasra}{Moral et~al.}{2006}]{delmoral2006SMCsampler}
Moral, P.~D., A.~Doucet, and A.~Jasra (2006).
\newblock Sequential {M}onte {C}arlo samplers.
\newblock {\em J.R. Statist. Soc. B\/}~{\em 68}, 411--436.

\bibitem[\protect\citeauthoryear{Neal}{Neal}{2001}]{neal2001annealedIS}
Neal, R.~M. (2001).
\newblock Annealed importance sampling.
\newblock {\em Statistics and Computing\/}~{\em 11}, 125--139.

\bibitem[\protect\citeauthoryear{Neal}{Neal}{2003}]{neal2003slice}
Neal, R.~M. (2003).
\newblock Slice sampling.
\newblock {\em The Annals of Statistics\/}~{\em 31\/}(3), 705--767.

\bibitem[\protect\citeauthoryear{Neklyudov, Welling, Egorov, and Vetrov}{Neklyudov et~al.}{2020}]{neklyudov2020involutivemcmc}
Neklyudov, K., M.~Welling, E.~Egorov, and D.~Vetrov (2020).
\newblock Involutive {MCMC}: a unifying framework.
\newblock In {\em 37th International Conference on Machine Learning}.

\bibitem[\protect\citeauthoryear{Pandolfi, Bartolucci, and Friel}{Pandolfi et~al.}{2014}]{pandolfi2014genMTMpap}
Pandolfi, S., F.~Bartolucci, and N.~Friel (2014).
\newblock A generalized multiple-try version of the reversible jump algorithm.
\newblock {\em Computational Statistics and Data Analysis\/}~{\em 72}, 298--314.

\bibitem[\protect\citeauthoryear{{R Core Team}}{{R Core Team}}{2022}]{r}
{R Core Team} (2022).
\newblock {\em R: A Language and Environment for Statistical Computing}.
\newblock Vienna, Austria: R Foundation for Statistical Computing.

\bibitem[\protect\citeauthoryear{Rubin}{Rubin}{1981}]{rubin1981eightschools}
Rubin, D. (1981).
\newblock Estimation in parallel randomized experiments.
\newblock {\em Journal of Educational Statistics\/}~{\em 6\/}(4), 377--401.

\bibitem[\protect\citeauthoryear{Salimans, Kingma, and Welling}{Salimans et~al.}{2019}]{salimans2019MCMCvariationalgap}
Salimans, T., D.~P. Kingma, and M.~Welling (2019).
\newblock Markov chain {M}onte {C}arlo and variational inference: bridging the gap.
\newblock In {\em International Conference on Machine Learning}.

\bibitem[\protect\citeauthoryear{Smith and {others}}{Smith and {others}}{2024}]{smithmcmcchains}
Smith, B.~J. and {others} (2024).
\newblock Mcmcchains.jl.

\bibitem[\protect\citeauthoryear{{Stan Development Team}}{{Stan Development Team}}{2024}]{stan2024}
{Stan Development Team} (2024).
\newblock {\em Stan Reference Manual}.

\bibitem[\protect\citeauthoryear{Surjanovic, Syed, Bouchard-C\^ot\'e, and Campbell}{Surjanovic et~al.}{2022}]{surjanovic2022variationalreference}
Surjanovic, N., S.~Syed, A.~Bouchard-C\^ot\'e, and T.~Campbell (2022).
\newblock Parallel tempering with a variational reference.
\newblock In {\em Advances in Neural Information Processing Systems}.

\bibitem[\protect\citeauthoryear{Tanner and Wong}{Tanner and Wong}{1987}]{tanner1987dataaugmentation}
Tanner, M.~A. and W.~H. Wong (1987).
\newblock The calculation of posterior distributions by data augmentation.
\newblock {\em Journal of the American Statistical Association\/}~{\em 82\/}(398).

\bibitem[\protect\citeauthoryear{Vehtari, Gelman, Simpson, Carpenter, and B\:urkner}{Vehtari et~al.}{2021}]{vehtari2021ranknormalization}
Vehtari, A., A.~Gelman, D.~Simpson, B.~Carpenter, and P.-C. B\:urkner (2021).
\newblock Rank-normalization, folding, and localization: An improved \^{R} for assessing convergence of {MCMC} (with discussion).
\newblock {\em Bayesian Analysis\/}~{\em 16\/}(2), 667--718.

\bibitem[\protect\citeauthoryear{Wang, Wang, and Bouchard-C\^{o}t\'{e}}{Wang et~al.}{2020}]{wang2020ASMCphylo}
Wang, L., S.~Wang, and A.~Bouchard-C\^{o}t\'{e} (2020).
\newblock An annealed sequential {M}onte {C}arlo method for {B}ayesian phylogenetics.
\newblock {\em Systematic Biology\/}~{\em 69\/}(1), 155--183.

\bibitem[\protect\citeauthoryear{Wang, Ge, Doig, and Wang}{Wang et~al.}{2021}]{wang2021semiparametricSMC}
Wang, S., S.~Ge, R.~Doig, and L.~Wang (2021).
\newblock Adaptive semiparametric {B}ayesian differential equations via sequential {M}onte {C}arlo.
\newblock {\em Journal of Computational and Graphical Statistics\/}~{\em 31\/}(2).

\bibitem[\protect\citeauthoryear{Wickham}{Wickham}{2016}]{ggplot2wickham2016}
Wickham, H. (2016).
\newblock {\em {ggplot2}: Elegant Graphics for Data Analysis}.
\newblock New York City, USA: Springer-Verlag New York.

\bibitem[\protect\citeauthoryear{Yang, Levi, Craiu, and Rosenthal}{Yang et~al.}{2019}]{yang2019componentwise}
Yang, J., E.~Levi, R.~Craiu, and J.~S. Rosenthal (2019).
\newblock Adaptive component-wise multiple-try {M}etropolis sampling.
\newblock {\em Journal of Computational and Graphical Statistics\/}~{\em 28\/}(2), 276--289.

\bibitem[\protect\citeauthoryear{Zhou, Johansen, and Aston}{Zhou et~al.}{2016}]{zhou2016toward}
Zhou, Y., A.~M. Johansen, and J.~A. Aston (2016).
\newblock Toward automatic model comparison: an adaptive sequential {M}onte {C}arlo approach.
\newblock {\em Journal of Computational and Graphical Statistics\/}~{\em 25\/}(3), 701--726.

\end{thebibliography}
\end{document}